\documentclass[lettersize,journal]{IEEEtran}
\usepackage{cite}
\usepackage{amssymb}
\usepackage{amsmath}
\usepackage{amsmath,amssymb,amsfonts}
\usepackage{algorithmic}
\usepackage{graphicx}
\usepackage{textcomp}
\usepackage[hyphens]{url}
\usepackage{xcolor, soul}

\sethlcolor{red}
\def\BibTeX{{\rm B\kern-.05em{\sc i\kern-.025em b}\kern-.08em
    T\kern-.1667em\lower.7ex\hbox{E}\kern-.125emX}}

\usepackage{amsmath, amssymb, graphics, setspace}
\newcommand{\mathsym}[1]{{}}
\newcommand{\unicode}[1]{{}}

\usepackage{caption}
\usepackage{subcaption}
\begin{document}

\title{Inter-Set Correlation-Aware Replacement Policy for Energy Efficient RTM-based Caches}
\author{Elham Cheshmikhani and Hamed Farbeh,~\IEEEmembership{Member,~IEEE}
\thanks{E. Cheshmikhani (Corresponding Author) is with the Department of Computer Science and Engineering, Shahid Beheshti University, Tehran, Iran. \\
H. Farbeh is with the Department of Computer Engineering, Amirkabir University of Technology, Tehran, Iran.}
\thanks{Manuscript received July 23, 2025.}}

\markboth{IEEE Transactions on Emerging Topics in Computing,~Vol.~xx, No.~x, August~2025}%
{Shell \MakeLowercase{\textit{et al.}}: A Sample Article Using IEEEtran.cls for IEEE Journals}


\maketitle

\begin{abstract}
Today's data-centric applications demand cache architectures that can scale with growing workloads while maintaining high performance and energy efficiency. 
Fundamental issues such as excessive area consumption and leakage power are increasingly challenging traditional SRAM-based caches, thereby motivating the exploration of non-volatile alternatives. 
Among these, \textit{racetrack memory} (RTM) stands out due to its remarkable storage density, achieved through nanowires hosting sequential magnetic domains that can be manipulated via domain wall or skyrmion techniques. 
Despite its advantages, racetrack memory's inherent serialized access introduces considerable shift overhead, leading to elevated energy consumption and latency. 
In this paper, we analyze the placement strategies in conventional RTM-based last-level caches and identify that current methods trigger redundant shift operations as a shift intended to read one block fails to preposition other blocks subsequently accessed. 
To resolve this, we introduce an innovative data placement and replacement scheme that intelligently groups correlated blocks, ensuring that a single shift not only retrieves the target block but also aligns subsequent blocks closer to the access port. 
Our simulation results using the gem5 simulator and the SPEC CPU2017 benchmarks reveal that our scheme reduces shift overhead by 49.0\% and cache energy consumption by 35.9\% with negligible performance impact.
In addition, this scheme exhibits robust scalability to longer nanowire tracks for higher cache density.

\end{abstract}


\section{Introduction}
In the era of data-intensive computing, on-chip caches have become indispensable components of modern multicore processors, playing a critical role in bridging the performance gap between high-speed cores and slower main memory. 
As applications grow in complexity and datasets expand exponentially in fields such as machine learning, big data analytics, and scientific simulations, the demand for larger on-chip caches has become increasingly pressing. 
Larger caches enable processors to store and access vast amounts of frequently used data, reducing memory latency and improving overall computational efficiency. 

For example, Intel's Xeon Scalable processors feature expansive shared last-level caches to enhance performance in data center workloads \cite{nassif2022sapphire}. AMD's EPYC processors are equipped with substantial cache sizes, enabling high efficiency in server applications with large data requirements \cite{
	bhargava2024amd}. Meanwhile, IBM’s Power Systems integrate extensive cache hierarchies to tackle computationally demanding tasks like deep learning and high-performance simulations \cite{IBM2021}.
These trends underscore the necessity of scaling cache sizes to meet the challenges posed by today's data-driven applications.

The growing need for larger last-level caches (LLCs) in multicore processors has exposed critical challenges associated with traditional SRAM technology \cite{hameed2023energy}. 
As cache sizes increase, SRAM struggles with scaling limitations, high static power consumption, and substantial area requirements, posing efficiency and cost challenges, especially in advanced nodes 
. 
To overcome these limitations, emerging non-volatile memory (NVM) technologies such as Resistive RAM (ReRAM), Spin-Transfer Torque Magnetic RAM (STT-MRAM), Ferroelectric RAM (FeRAM), and Racetrack Memory (RTM) have gained significant attention \cite{
	sun2024roadmap, khan2023downshift, hadizadeh2021copa, 
	 bera2024spimulator, cheshmikhani2019robin}. 
Among these, RTM stands out for its exceptionally high density, which far exceeds that of other NVMs, making it a promising solution for future cache architectures \cite{cheshmikhani2024low, sharma2017nonvolatile, archer2020foosball, venkat2023magnetic, bereholschi2024evaluating}. 
In addition to its density, RTM's ability to provide low latency and high endurance adds to its appeal \cite{cheshmikhani2024low, li2023ultralow}.

%


Racetrack Memory is built on a unique structure where each memory cell consists of a nanowire as the storage element, embedded with a series of multiple magnetic domains \cite{an2024streampim, abdullah2024enhanced, kumar2022domain}. 
These domains represent binary data, and the position of each domain within the nanowire is manipulated using electrical currents, a process known as domain wall motion \cite{sima2023correcting, kumar2023ultralow}. 
Domain wall technology leverages the boundaries between magnetic regions to encode and shift data efficiently, enabling high-density storage. 
In addition to domain wall memory (DW-RTM or DWM), skyrmion-based racetrack memory (SK-RTM or SKM) has emerged as a promising alternative. 
Skyrmions are nanoscale, stable magnetic structures that require a lower current to manipulate compared to domain walls, thereby offering energy efficiency and scalability advantages \cite{karino2024domain, yang2024toward, nishitani2023connected, kang2019comparative, chen2018process, al2023multi}. 
To access data in racetrack memory, an access port is employed to read the magnetic states serially as the domains or skyrmions are shifted past it. 
However, the serial nature of data access poses challenges, introducing energy and latency overheads, particularly for data stored farther from the access port. 
Recent reports show that the shift operation accounts for more than 50\% of cache energy consumption \cite{sun2013cross, chen2018process}
These trade-offs highlight the ongoing efforts to optimize both domain wall and skyrmion technologies for next-generation cache systems.

Several recent studies have focused on addressing the high energy consumption and variable access latency in RTM caused by its serialized data access. 
One promising approach involves optimizing data placement on software-managed RTMs, such as main memory and scratchpads, to minimize unnecessary shifts and reduce energy overhead \cite{khan2023downshift, hui2023optimizing}. 
Modifying cache replacement policies to account for shift costs during victim selection on a cache miss has also been explored to enhance efficiency. 
Another strategy includes increasing the number of access ports per track while optimizing the access transistor size to balance performance and area requirements. 
Utilizing non-uniform tracks for different cache ways and adaptively resizing track lengths based on workload characteristics have also shown potential in reducing shift-related delays. 
Additionally, data compression techniques have been investigated to lower shift overheads by effectively reducing the amount of data to be accessed. 
These efforts collectively aim to mitigate the challenges associated with serialized access in RTM and improve its viability as a high-performance memory solution in modern systems.

These solutions also come with notable drawbacks that hinder their practical adoption. 
Firstly, software-based optimizations are not directly applicable to hardware-managed on-chip caches, limiting their utility in traditional cache hierarchies. 
Secondly, increasing the number of access ports or shrinking the nanowire length contradicts the racetrack memory's primary advantage—its exceptional density—thereby reducing the fundamental benefit of this technology. 
Thirdly, employing non-uniform cache structures to map blocks based on access patterns complicates cache design and management while providing only limited efficiency gains. 
Lastly, modifying cache replacement policies to trade off the shift cost and block usefulness often results in evicting frequently used blocks, thereby degrading cache performance. 
To fully harness the potential of racetrack memory in LLCs and address these limitations, it is essential to rethink the cache architecture and redesign cache management and replacement policies to align with the unique characteristics of RTM.

In this paper, we address the inefficiencies of data placement in conventional RTM-based LLCs that lead to excessive shift operations. 
Our investigation reveals that the current data placement strategy is unsuitable for serialized access, as a shift operation to access one block does not benefit subsequent accesses to other blocks residing on the same track. 
To mitigate this issue, we propose a novel data placement and replacement policy designed to optimize the mapping of cache blocks to track positions. 
This policy ensures that a shift operation to access a block also positions subsequently accessed blocks closer to the access port, thereby reducing shift overhead. 
To achieve this, we first analyze cache access patterns and detect a high probability of correlation in consecutive accesses to cache sets, identifying blocks that are frequently accessed together. 
Based on these findings, we introduce a mapping strategy where cache sets and ways are organized in such a manner that correlated blocks have a higher likelihood of residing on the same track. 
Additionally, our proposed replacement policy considers both the correlation of an incoming block with neighboring blocks and the recency history of victim candidates, effectively balancing block usefulness and shift cost to enhance overall cache performance.

Our experimental evaluation, conducted using the gem5 simulator \cite {binkert2011gem5} and the SPEC CPU2017 benchmark suite \cite{speccpu2017}, demonstrates the effectiveness of the proposed scheme. 
The results show that our scheme reduces the shift cost by 49.0\% compared to state-of-the-art methods, significantly alleviating one of the key limitations of RTM. 
This reduction in shift operations translates into a 35.9\% decrease in cache energy consumption, making the proposed scheme highly energy-efficient. 
Furthermore, the scheme exhibits strong scalability, maintaining its advantages even with larger track lengths, highlighting its suitability for next-generation racetrack-based last-level caches.

The remainder of this paper is organized as follows. Section 2 provides an overview of the related work, highlighting previous efforts and approaches to addressing the challenges of RTM in various memory levels. Section 3 presents our observations and motivation, followed by a detailed description of the proposed data placement and replacement scheme aimed at reducing shift operations in RTM-based LLC. In Section 4, we evaluate the effectiveness of our scheme, reporting results that demonstrate its significant advantages in reducing shift cost, energy consumption, and improving scalability and performance. Finally, Section 5 concludes the paper by summarizing our contributions and discussing potential directions for future research in the field of racetrack memory optimizations.



%

\section{Related Work}
Several recent studies addressed the high energy consumption and variable access latency due to its serialized access.
Optimizing data placement on software-managed RTMs, e.g., main memory and scratchpad, for shift reduction; modifying the cache replacement policy for taking the shift cost into account on a cache miss for victim selection; increasing the number of access ports per track and optimizing the access transistor size; utilizing non-uniform tracks for cache ways; adaptively resizing the track length; and data compression are among the main efforts for reducing the shift cost and enhancing the RTM-cache performance.
In the following, we explore these approaches in more detail.

TapeCache addressed the variable access latency and high shift cost of RTM-caches and proposed a combination of device, circuit, and architectural schemes to mitigate them \cite{venkatesan2015cache}. 
A new write mechanism was proposed at the device level based on track shift to reduce the write latency and energy consumption.
At the circuit level, DWM cells were tailored to the different cache level requirements, where fast single-bit DWM cells were suggested for high-speed L1 caches, and high-density multi-bit DWM cells were employed in L2 caches.
To mitigate the shift cost in lower cache levels utilizing multi-bit cells, the architectural solution of TapeCache is to use multi read-only ports beside the main read-write port in the nanowires.
In addition, a fraction of cache ways are made up of single-bit cells for fast access to store the frequently accessed cache blocks.

A circuit-level design space exploration was conducted in \cite{motaman2014synergistic} to find the proper number of access ports per nanowire for trading off the access latency, fraction of functional bits per cell, and cell density.
This work also studied the access transistor sizing for optimizing the area and latency, considering the read disturb and write latency.
The cache ways are non-uniform in terms of shift speed by manipulating the shift current, and it tries to assign fast ways to more active cache blocks and slow ways to lower-frequently accessed blocks.

Considering multi-port RTM caches and bit-interleaved data mapping, it was suggested by \cite{larimi2016power} that instead of track-based interleaving, where each bit or block is stored in a separate track, data bits are stored in a port-based interleaving manner.
In this case, the number of bits of a block stored on a track is proportional to the number of ports per track.

A design space exploration was conducted in \cite{sun2014design} for access port organization in RTM-LLCs assuming large enough write-read ports and small read ports.
Based on the observations, this work proposed a mixed array organization for cache ways where a fraction of ways is read-optimized using a single large enough write-read port and multiple small read ports, and other ways use multiple uniform write-read access ports.
The read-intensive blocks, e.g., instruction blocks, are mapped to read-optimized ways, and heavily updated blocks are mapped to uniform ways.
Meanwhile, underutilized cache ways are disabled dynamically to shrink the cache associativity for faster data access by reducing the number of shifts.

DyReCTape is a reconfigurable RTM-cache that dynamically inactivates a portion of track bits to reduce the access latency at the cost of reducing cache capacity \cite{ranjan2015dyrectape}.
This scheme monitors the application's sensitivity to shift latency and cache capacity and switches between the two cache modes.

The impact of process variations (PVs) on the performance of SKM-cache was addressed in \cite{chen2018process}, and a PV-aware data management scheme was proposed.
This work showed that PVs can change the shift and read access latency of cache lines.
Based on PV analysis, the proposed scheme partitions the cache banks into several parts with different access latencies and dynamically swaps data between them based on their access pattern.

The high write energy consumption of Skyrmion-based RTM-cache was addressed in \cite{yang2022evolving}, and an adaptive write mode strategy was proposed that switches between word-based and bit-interleaved mapping architectures to provide high access performance during cache fetches and low energy consumption during cache evictions.
This work targeted LLC.

Instead of leaving the access port on the last access data, a prediction technique was proposed in \cite{colaso2019architecting} to place the access port on the next predicted data block to reduce the shift latency on a data request.
This technique tries to predict the length and direction of the next port displacement and preshifts the track for hiding the shift latency.
The proposed preshifting technique is inspired by the existing correlation-based prefetchings.

Data compression techniques were utilized by \cite{xu2015multilane} to reduce the access latency and shift energy in RTM-LLC.
On an access to a compressed cache block, only the tracks containing valid data bits are shifted using an independent shifting mechanism instead of simultaneous shifting of all tracks in a \textit{Domain~Block~Cluster} (DBC).
This scheme exploits a skewed alignment technique for storing compressed blocks in different non-overlapping offsets of cache blocks.

Targeting embedded applications, a compiler-based data placement technique was presented in \cite{chen2015optimizing, chen2016efficient} for reducing the number of shift operations.
This work proposes a grouping-based data placement heuristic algorithm for extracting the memory address correlation from data access patterns.
Consecutively accessed data items are mapped to different DBC segments with the same offset for minimizing the shift operations.

An integer linear programming (ILP) formulation was presented in \cite{khan2019shiftsreduce, khan2020generalized} for data placement in RTMs.
Using ILP, this work tries to heuristically compute the memory offsets considering the temporal locality of accesses via profiling the application behavior. 
The schemes in \cite{khan2019shiftsreduce, khan2020generalized}, \cite{chen2015optimizing, chen2016efficient} are inapplicable to general-purpose systems and RTM-caches. 

Fast-Track Cache is a new data array organization proposed for RTM-based L1 data cache \cite{tarrega2022fast}.
This scheme arranges cache lines in a header-interleaved manner to leverage the spatial and temporal locality for shift reduction.
To enhance the system performance, this scheme merges L1 data and L2 caches into a larger L1 data cache.

Tag array in RTM caches is conventionally implemented by SRAM or single-bit RTM cells to avoid the shift cost.
Assuming multi-bit cells in a tag array, BlendeCache \cite{hameed2022blendcache} suggested a cache structure capable of parallelizing tag shift with data block shift to hide the tag access latency.
The cache loses the benefits of its associative structure in this scheme as BlendCache forces the cache to be in a direct-mapped structure.

Hameed et al. \cite{hameed2023energy} proposed a Shift-Aware replacement policy  (SAR) that reduces block shift costs on a cache miss.
This policy selects victims by balancing block recency with their distance to the access port, evicting the rarely used block with the lowest shift cost instead of the oldest block as in LRU.
While SAR reduces shift costs during replacement, it offers no benefit to blocks already residing and being accessed in the cache.


\section{Proposed Method}
\label{prop}
The main criteria in RTM-based caches is minimizing the number of track shifts to enhance the performance and energy saving without impacting the cache miss rate.
This section first presents our observations to illustrate the deficiency of existing RTM cache structures and managements as the motivation of this work. 
Then, the proposed cache structure is introduced, which can reduce the number of shifts based on the observations in a well-managed cache.
Finally, our cache replacement policy is presented.
\subsection{Observations and Motivation} 
\label{motiv}
\begin{figure*}[t]
	\centering
	\includegraphics[width=0.61\linewidth]{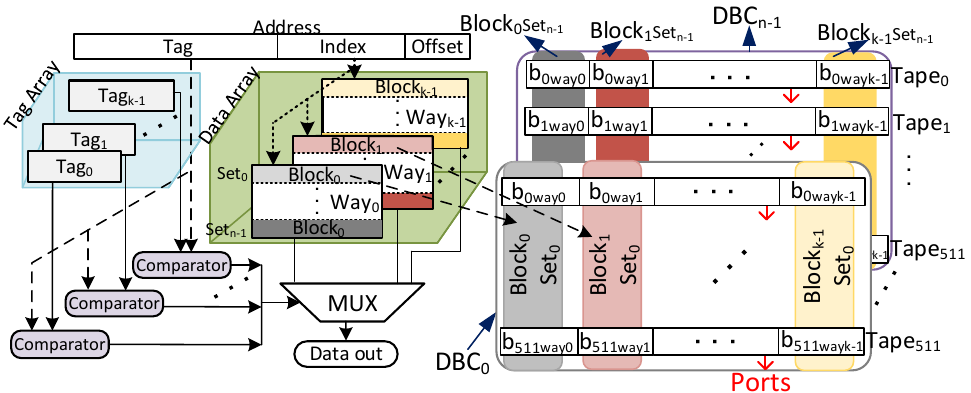}\vspace{-5pt}
	\caption{RTM-cache structure in conventional design, where each cache set (all its ways) is constructed by a DBC consisting of 512 parallel k-bit tapes.}\vspace{-5pt}
	\label {sheklejadid}
\end{figure*}

In conventional RTM-cache, each DBC constructs a cache set, where data blocks in a set of k-way associative cache share 512 tracks (or tapes) of length K (assuming a cache block size of 64 bytes).
In other words, each K-bit track in a DBC contains a single bit of K ways in the set. 
Therefore, the access port for each set is always positioned on the way containing the MRU block in the set, and for new access to the set, all 512 tracks are shifted by the distance of the way containing the MRU block and the currently requested way. 
This means that for each access to a cache set, all ways in the set are shifted alongside the requested way.
This structure is depicted in Fig. \ref{sheklejadid}.

\begin{figure*}[t]
	\centering
	\includegraphics[width=0.75\linewidth]{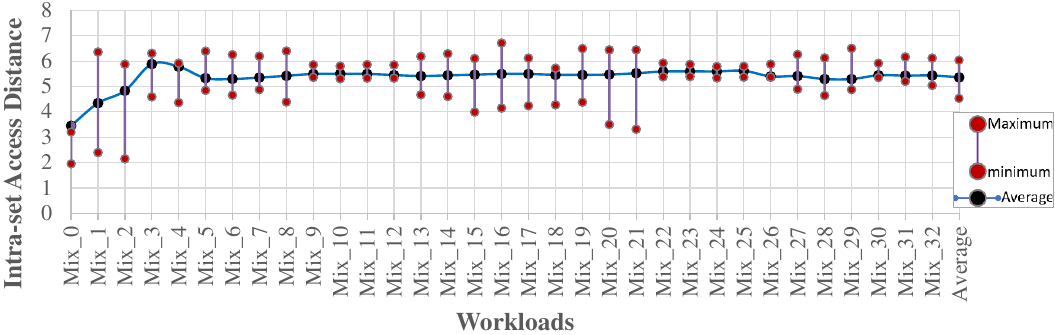}
	\caption{Intra-set distance in consecutive accesses to LLC sets in various workloads. Intra-set distance value determines the number of shifts requires for each access to a set since the most recently used (MRU) block in each set is alwayes on the access port.}
	\label {per-set-dis}
\end{figure*}

The deficiency of this RTM-cache structure is two-fold: 1) there is no direct relationship between the content of cache lines in a set and the ways adjacency; 2) a cache set contains the most unrelated (distant) memory blocks according to the set-associative cache placement policy.
Therefore, though by an access to a cache block, all K-1 ways in a set are shifted in favor of a single way containing the requested data, they do not benefit from this shift for the subsequent accesses to that set.
Owing to this fact, the number of shifts per access is expected to be random due to unrelated consecutive accesses and block content.

Fig. \ref{per-set-dis} depicts the number of shifts required per LLC access in a 16-way set-associative RTM-cache running different combinations of workloads from the SPEC CPU2017 benchmark suite \cite{speccpu2017} on a quad-core processor. 
The details of system configuration are given in Section 5 (Table \ref{table}).
The figure depicts that the average number of shifts is 5.4 bits. 
The lowest and highest values for the number of shifts are observed in Mix1 and Mix3, which are 3.5 and 5.9 bits, respectively. 
As different sets of cache experience various values for the number of shifts based on the workloads' memory access pattern, Fig. \ref{per-set-dis} also shows the minimum and maximum values beside the average for each workload. 
The results show that the diversity in the number of shifts in cache sets is low in all workloads, which confirms the randomness in accessing the ways of all cache sets. 
It should be noted that, although the average number of shifts for a 16-way cache may be expected to be 7.5 bits, the higher frequency of accesses to MRU blocks (with no need to shift) deviates the average value.
Hence, by randomness in the number of shifts per set, we are referring to accessing those blocks that are not located on the access port.

\begin{figure*}[t]
	\centering
	\includegraphics[width=0.61\linewidth]{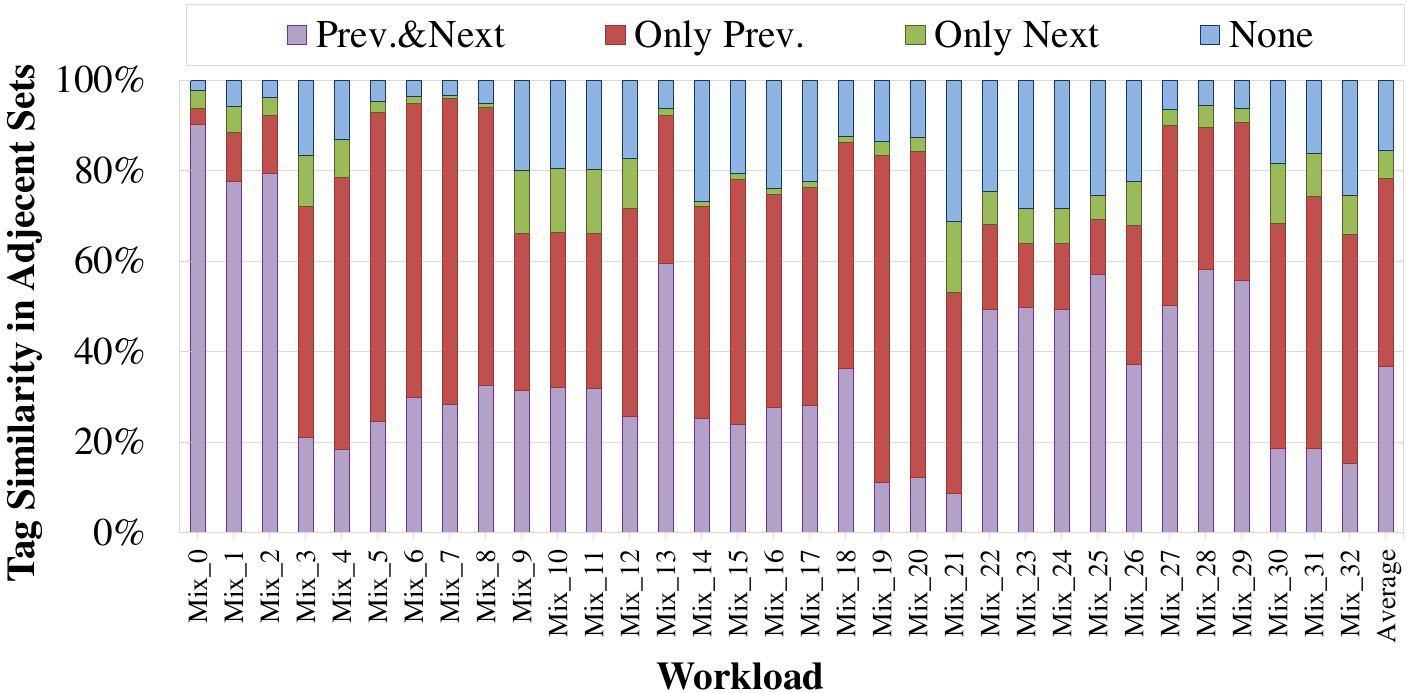}\vspace{-5pt}
	\caption{Probability of finding a tag in adjacet upper and lower sets (previous and next sets) similar to the tag of requested blocks on a cache miss occurrence in a set. Similar tags based on our definition belong to correlated blocks since they are located in neighboring blocks in main memory.}\vspace{-5pt}
	\label {HF-BrfStat}
\end{figure*}


\begin{figure*}[t]
	\centering
	\subfloat[]{\includegraphics[width=0.78\linewidth]{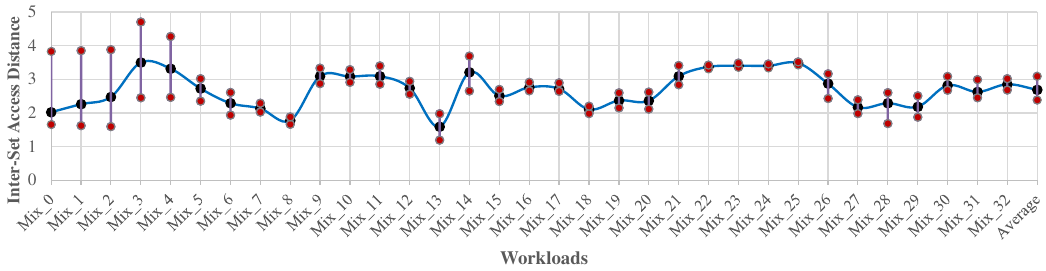}}\vspace{-1pt}
	\hspace{25pt}
	\subfloat[]{\includegraphics[width=0.78\linewidth]{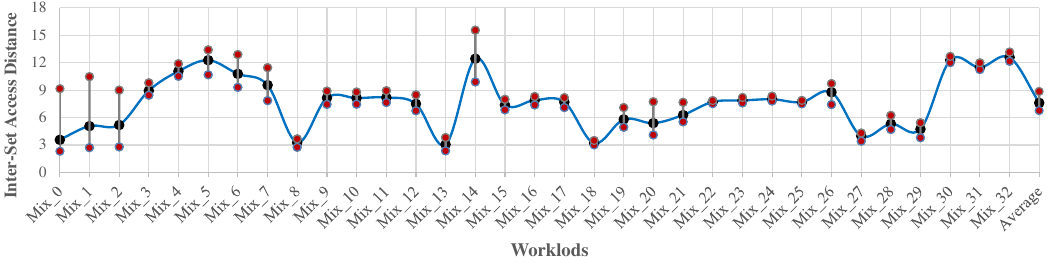}}\vspace{-3pt}
	\caption{Inter-set distance in consecutive accesses to LLC in various workloads for a) Supersets with 16 sets and b) Supersets with 64 sets. Inter-set distance value for supersets with minimum and maximum distances are depicted alongside the average value of all supersets per workload, illustrating a low variance in inter-set distances among cache supersets.}\vspace{-5pt}
	\label{fig:basics}
\end{figure*}

\begin{figure*}[t]
	\centering
	\includegraphics[width=1\linewidth]{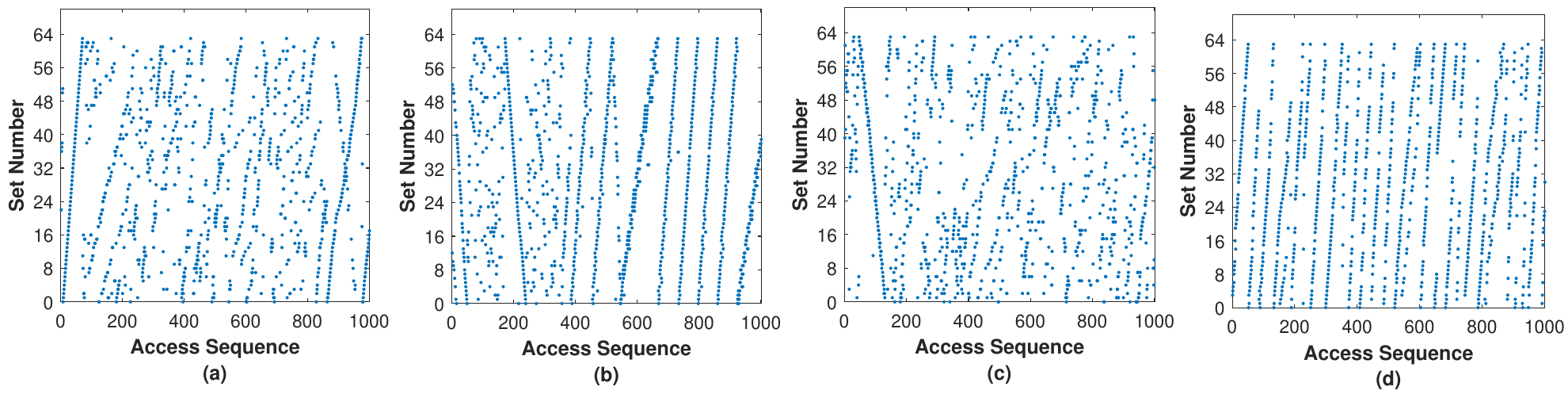}\vspace{-5pt}
	\caption{Set number accessed in consecutive requests to a randomly-selected superset of RTM-cache for a window of 1000 accesses in four randomly-selected workloads. Low distance between two consecutively-accessed sets depicts correlation between accessing to neighboring sets.}\vspace{-10pt}
	\label {matlab dot figure}
\end{figure*}

As the shift operation for accessing a block in the cache inevitably shifts all other K-1 blocks sharing the same tracks in a DBC, this operation should ideally bring blocks that will be accessed in the subsequent requests near to the access port.
The realization of this dream requires 1) mapping the most correlated cache blocks to the same DBC to share tracks shifted simultaneously and 2) placing these correlated blocks in the track positions in order according to the sequence of accesses to them. 
If doing so, when shifting the track to place the requested block on the access port, the blocks moving toward the port are those that will be requested in the subsequent accesses.
Hence, the minimum number of shifts is incurred for each access.

The key questions are: 1) How to find the most correlated cache blocks, 2) how to map them onto the same DBC without violating the cache placement policy, and 3) how to place the order of blocks sharing a track according to their order of access.
To answer these questions, we need to take a closer look at cache placement policy and the mapping of memory blocks into a set associative cache.
Adjacent blocks in the main memory have consecutive addresses, which is reflected in the \textit{Index} part of the block address. 
These blocks that are the most correlated due to spatial locality in accessing the memory locations are mapped into adjacent cache sets.
Therefore, the most correlated memory blocks are spread out among the adjacent sets, while the competing blocks inside a set are the most unrelated ones, as they are the most distant blocks in the main memory.
The former blocks are similar in their \textit{Tag} part and differ in their \textit{Index}, and the latter are in reverse.
Therefore, it is anticipated that the adjacent cache sets contain blocks with similar tags and a large fraction of consecutive accesses to the cache are to adjacent sets due to spatial locality.

We conducted a set of experiments to investigate the above anticipation.
Fig. \ref{HF-BrfStat} depicts the probability of finding similar tags in adjacent cache sets on accessing a cache block for various workloads. 
The results show that for 84.5\% of accesses to cache blocks in a set, there is a block in at least one of the two adjacent sets with a similar tag. 
For 36.9\% of blocks, the similar tag is found in sets on both sides, and for 41.5\% and 6.1\% of blocks, it is found in either the upper or lower set, respectively. 
This observation confirms the residency of correlated blocks in adjacent cache sets.

To investigate the locality in accesses into adjacent sets, we measure the inter-set distances of consecutive accesses into LLC. 
To have a meaningful measurement, the cache sets are assumed to be grouped, and the inter-set distances inside each group, the so-called \textit{superset}, are considered. In other words, the cache is partitioned into supersets of equal size, and the distances of accesses per superset are taken into account.
Fig. \ref{fig:basics} shows the average inter-set distances of different workloads for supersets of size 16 and 64. 
The results depict that the average distance of two consecutive accesses is only 2.7 and 6.8 in supersets containing 16 and 64 sets, respectively. 
Comparing the intra-set distances for accessing consecutive cache ways inside a set observed in Fig. \ref{per-set-dis}, a tight correlation in accessing adjacent sets is evident.
The figure also depicts the value for supersets with minimum and maximum inter-set distances besides the average values.
The results show a low variation in inter-set distances among both 16- and 64-set supersets for all workloads.

\begin{figure}[t]
	\centering
	\includegraphics[width=1\linewidth]{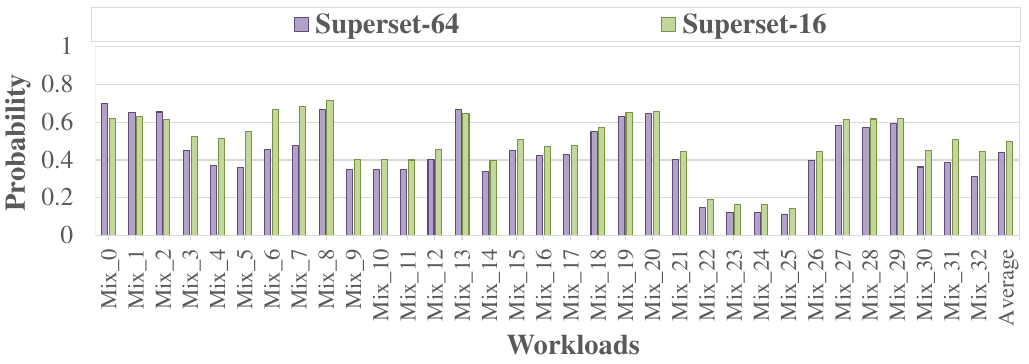}
	\caption{Probability of accessing adjacent sets (inter-set distance = 1) in consecutive accesses to a superset in 16- and 64-set supersets for various workloads.}
	\label {shekl2.5}
\end{figure}

Fig. \ref{matlab dot figure} illustrates a more detailed view of access sequences to cache sets.
As an exemplary pattern, a randomly-selected 1000 accesses to a superset in four workloads is depicted.
The dots in the figure show the set number referred to (vertical axis) in each access (horizontal axis).
As can be seen, the inter-set distances are dominated by a sequential access pattern where sets with short distances are accessed consecutively.
More precisely, Fig. \ref{shekl2.5} shows the fraction of accesses with the inter-set distance of exactly 1.
The result demonstrates that 49.5\% and 43.9\% of accesses are to adjacent sets in 16-set super-sets and 64-set supersets, respectively.

\begin{figure}[t]
	\centering
	\includegraphics[width=1\linewidth]{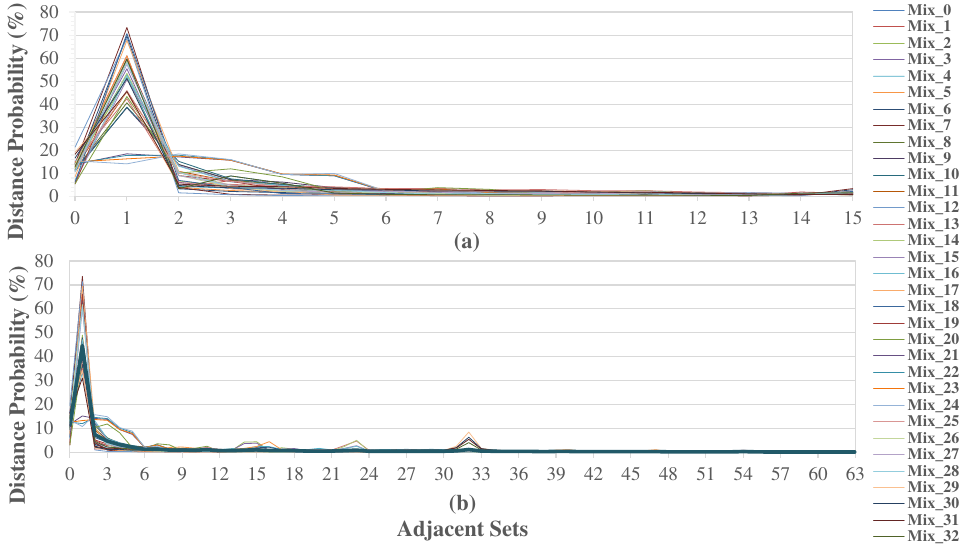}
	\caption{Probabilities of all inter-set distances in accessing supersets for various workloads considering 16- and 64-set supersets for RTM-cache.}
	\label {distance-of-consecuitive sets}
\end{figure}

More detailed results on the inter-set distance are given in Fig. \ref{distance-of-consecuitive sets}.
Considering all workloads, the contribution of all inter-set distances in accessing super-sets is illustrated. 
Besides the dominant contribution of inter-set distance of 1 (accessing adjacent sets), the contribution of inter-set distance of 0 is also considerable.
This result shows that the probability of accessing the most recently accessed set or one of its two adjacent sets is as high as 62.4\% and 55.7\% in 16- and 64-set supersets, respectively. 

\subsection{Proposed Cache Structure}
Based on the above observations and analysis, placing correlated blocks on the same DBCs is an opportunity to minimize the number of shifts. 
This means that instead of sharing each DBC between ways of each set, i.e., conventional structure, we call it \textit{horizontal DBC} hereafter, multiple consecutive sets share a DBC for placement of one of their ways.
Therefore, considering a 512-track K-bit DBC that constructs K blocks of a set in a conventional K-way associative cache, we propose to use this DBC for constructing way{i} of set{n} - set{n+k}, the so-called~\textit{vertical~DBC} structure.
In this manner, K DBCs construct K ways of a superset containing multiple sets (equal to track length) instead of constructing K sets in a K-way associative cache.

\begin{figure}[t]
\centering
\includegraphics[width=1\linewidth]{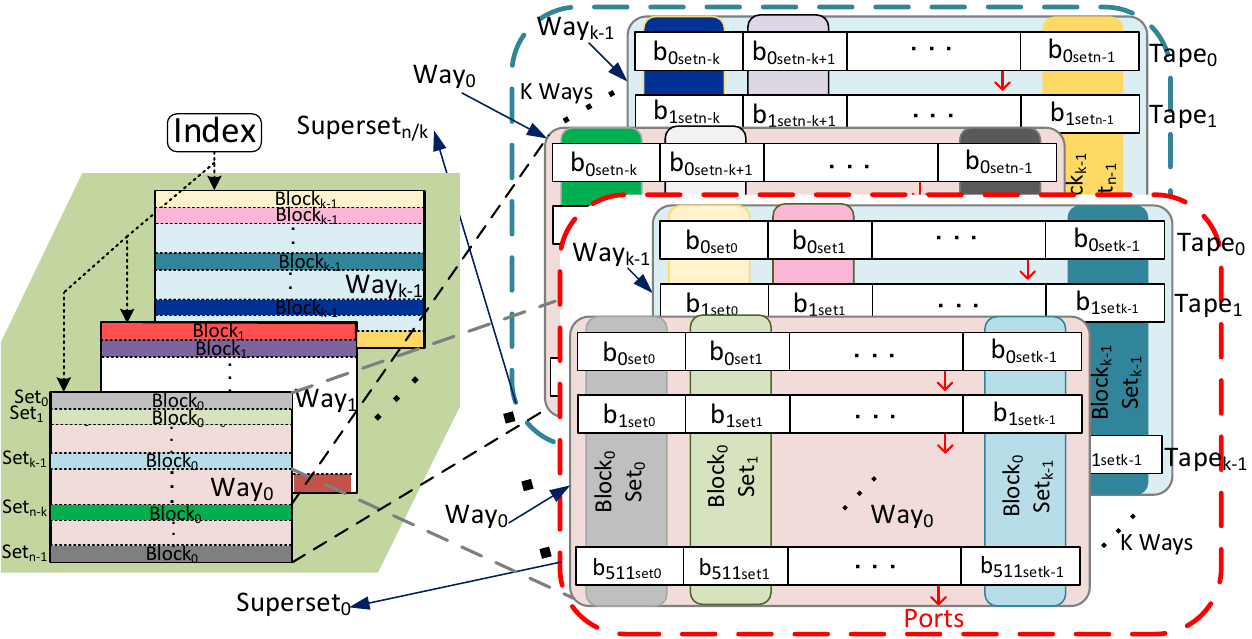}
\caption{Cache structure based on our observed inter-set access correlation.}
\label {arch}
\end{figure}

This structure is depicted in Fig. \ref{arch}. 
When accessing a cache block, only the containing way shifts to the access port, leaving other ways intact within the set.
Instead, the corresponding way in other sets in the superset is shifted accordingly.
In addition to the opportunity of placing correlated blocks on the same DBC, the length of tracks in vertical-DBC is not limited to the cache associativity and can be arbitrarily adjusted, which is not the case for conventional horizontal DBC.

A key question is how effectively correlated blocks are placed on the same vertical DBCs. Our observations indicate that correlated blocks tend to map to consecutive cache sets, allowing them to share a DBC and benefit from shifts during access to adjacent sets. 
However, there is an architectural challenge that should be explored and resolved.
While sharing DBCs between cache sets in a vertical structure \textit{potentially} places the correlated blocks on the same track, each set is constructed by multiple independently-managed DBCs, where each way in the set has its own DBC.

Correlated blocks only benefit from vertical DBCs if they reside on the same DBC (the same way) in adjacent sets. However, since each k-way associative superset shares k independent DBCs, an incoming data block can reside in any way, irrespective of its correlated blocks' location in adjacent sets. The cache replacement policy independently determines the location of each data block within the ways of each set. Consequently, a block allocated to $way_{i}$ in one set does not guarantee correlated data in an adjacent set will also be allocated to $way_{i}$. This can lead to correlated blocks being spread across different DBCs, negating the benefits of the vertical DBC structure.

\begin{figure}[t]
	\centering
	\includegraphics[width=1\linewidth]{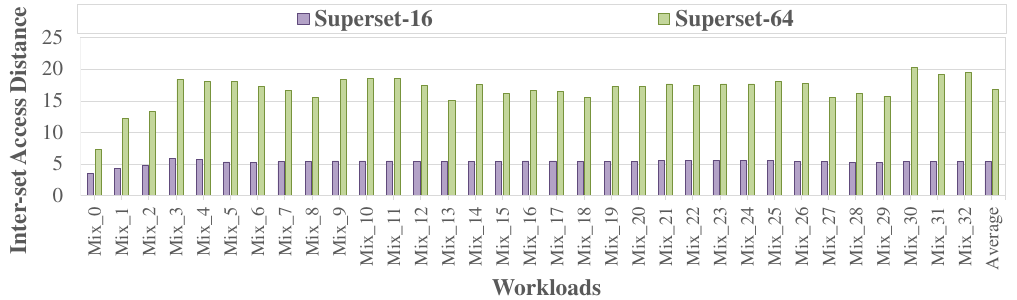}
	\caption{Number of shifts per access in the proposed vertical-DBC address mapping using conventional LRU replacement policy. The results show that conventional replacement policies cannot take advantage of inter-set access correlation.}
	\label {shekl9}
\end{figure}

Fig. \ref{shekl9} depicts the average number of shifts required per cache access in a vertical DBC structure of lengths 16 and 64 for different workloads.
As shown, an average of 5.4 and 16.8 shifts is required for each access in DBCs of length 16 and 64, respectively, although we have already observed the inter-set access distance of 2.7 and 6.8 in Fig. \ref{fig:basics}. 
This observation confirms the independent way allocation by cache replacement policy in adjacent sets, which causes randomness in correlated blocks' residency on DBCs.

\subsection{Proposed Cache Replacement Policy}
Replacing SRAM with RTM for LLC necessitates a cache redesign, encompassing both organizational restructuring and revised management policies, including replacement strategies. Conventional replacement policies treat block allocation within each set independently.
Existing replacement policies do not differentiate between ways inside a set based on their physical location but on their access sequence.
Hence, a new replacement policy is needed to take into account the RTM-cache organization and DBC structure constructing the cache sets and ways besides the recency of accesses and discovering temporal locality.

\begin{figure}[t]
	\centering
	\includegraphics[width=0.85\linewidth]{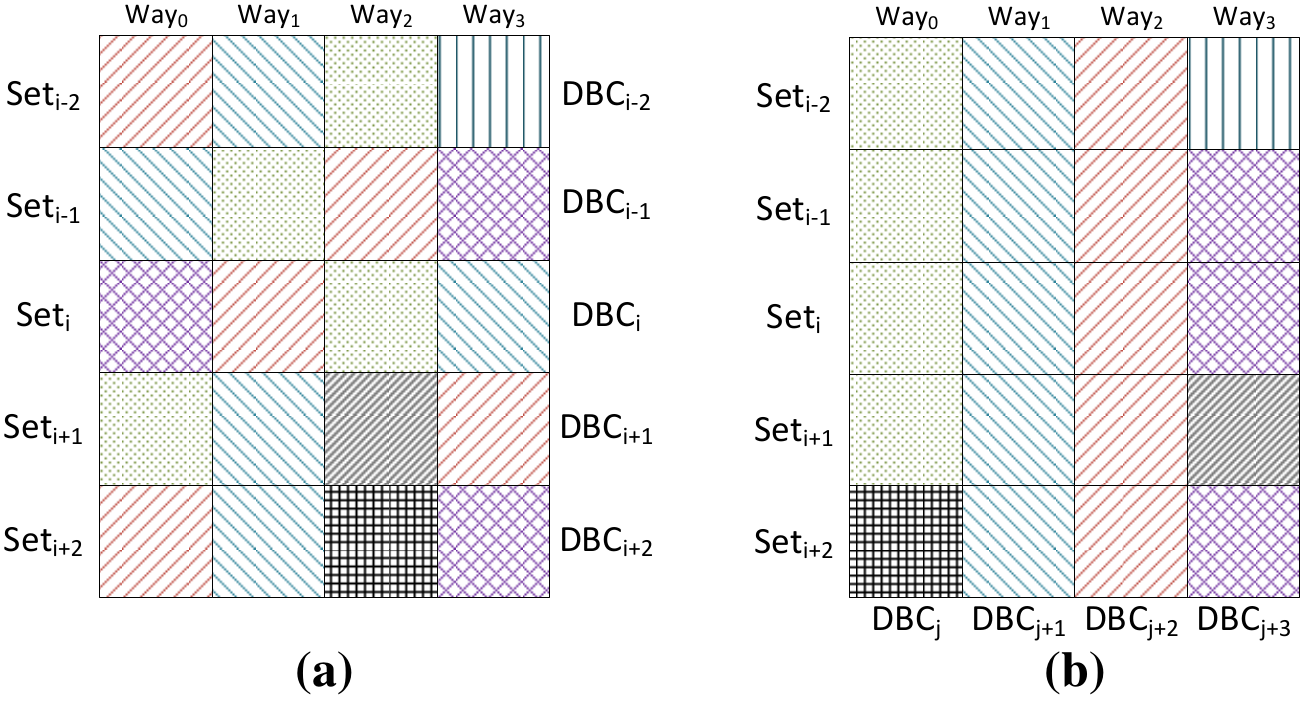}
	\caption{Visualization of cache block allocation in conventional replacement policy and ideal allocation based on the proposed vertical-DBC, where correlated blocks (depicted in similar patterns in neighboring sets) are mapped into same way number in the sets.}
	\label {imagine}
\end{figure}

Our proposed policy, named \textit{Inter-Set Correlation Aware} (ISCA) replacement policy, allocates cache blocks to incoming data on a cache miss in such a manner that the correlated blocks reside in the same cache way, i.e., the same DBC.
To this aim, ISCA, in its basic and simplest configuration, searches for a correlated block adjacent to sets on a cache miss, and the victim block is the one located in a way in which the correlated block is found.
Fig. \ref{imagine} visualizes an imaginary scenario of block allocation in five consecutive cache sets in 4-way associative configuration for conventional replacement policy (Fig. \ref{imagine}-(a)) and our favored allocation (Fig. \ref{imagine}-(b)).
As depicted in Fig. \ref{imagine}-(a), while there is a high probability of the existence of correlated blocks in adjacent sets (blocks in the same patterns are correlated), the placement of blocks in each set is independent of others, resulting in random allocation of ways to the blocks.
As it is highly probable to request correlated blocks in adjacent sets in consecutive accesses to a superset, an ideal block allocation is to place the correlated blocks in the same DBC (the same way in neighboring sets), as depicted in Fig. \ref{imagine}-(b).
Our replacement policy selects the victim blocks on cache misses in a manner that the incoming block is mapped to the way that its correlated blocks reside in its adjacent sets.

More precisely, ISCA answers the following questions:

\textbf{How to detect the correlated blocks?}
According to our observations and analysis in Section \ref{motiv}, correlated blocks are those located in neighboring memory locations and have similar \textit{Tags} in adjacent sets.

\textbf{How to find a correlated block in adjacent sets?}
Similar to finding a match in the referred set, the cache controller is modified to perform the tag comparison in its adjacent sets on a cache miss.

\textbf{Which adjacent set is searched first?}
As observed in Fig. \ref{HF-BrfStat}, the probability of finding a similar tag in the upper set is 41.5\%, in the lower set is 6.9\%, and in both sides is 36.9\%. 
Therefore, finding a similar tag in the upper set is significantly more probable, and it is searched for first.

\textbf{What to do if no correlated block is found?}
In this case, the victim block is the one whose way has the shortest distance to the access port.

\textbf{What is the performance overhead of searching adjacent sets?}
The search operation is only performed on cache misses and is overlapped with fetching the missed block from a lower memory level on a read miss, which lasts several times longer than the search operation.
On a writeback miss, on the other hand, the writeback operation is performed in the background, and the delay of the search operation is hidden.
Hence, the delay for extra tag comparison in adjacent sets is not in the critical path and is ignorable.

\textbf{What is the performance overhead of finding a way with the nearest access port when no correlated block is found?}
In parallel with the tag comparison operation on every access to a cache set (regardless of hit or miss occurrence), the distance of all access ports to the set is calculated to initiate the shift operation immediately after the detection of a hit.
The only modification required is to find the access port with minimum distance after detection of a cache miss, which is again overlapped with fetching the missed block from a lower memory level or performing writeback in the background.
Therefore, this operation imposes no performance overhead.

\subsection{ISCA Optimization}
Focusing on minimizing the number of shifts and allocating cache blocks only based on their correlation or distance to access ports by ISCA while ignoring the recency and temporal locality of blocks may increase the cache miss rate and degrade the performance.
To strike a balance between the number of shifts and the miss rate, ISCA can be integrated with the existing replacement policies, which focus on victim selection within single sets.
In other words, ISCA can exploit the existing replacement policies, e.g., LRU, RRIP, and SHiP, as an underlying scheme to filter the candidate blocks for eviction.

Instead of taking into account all blocks in a set, the underlying scheme determines a subset of more suitable blocks for eviction, and ISCA searches for correlated blocks in only the corresponding cache ways in the adjacent sets.
For example, considering LRU as the underlying replacement policy, ISCA can take into account the upper half/quarter of older blocks in a set for eviction candidates.
If no correlated block is found, only blocks in this subset are considered by ISCA for finding the block with the nearest access port.

\begin{figure}[t]
\centering
\includegraphics[width=1\linewidth]{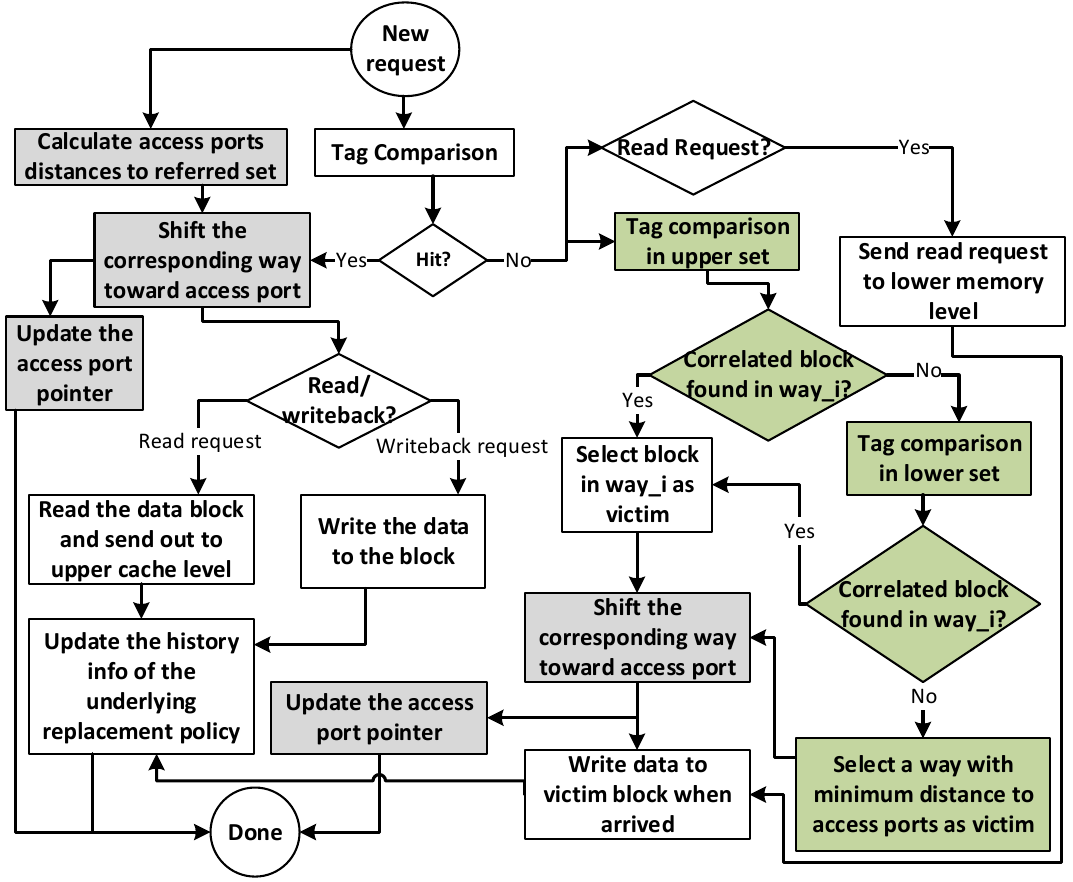}
\caption{Control flow of the cache in the proposed replacement policy. The boxes in green are the operations added by the proposed policy, the gray boxes are specific to all RTM-caches, and the white boxes the basic operations for all cache independent of memort technology.}
\label {flowchart}
\end{figure}

\subsection{ISCA Operational Details}
The flowchart in Fig. \ref{flowchart} depicts the detail of ISCA operations for a cache access.
After decoding the address \textit{Index} field and finding the referred set and its corresponding superset (using the upper part of \textit{Index}), the operations of tag comparison in the set and calculating the superset access port distances to the set are performed simultaneously.
On a hit, the cache way in the superset containing the requested block is shifted according to its distance to its corresponding access port. 
The data is read and sent out to the upper cache level for a read request or written to the block on a writeback request from a higher cache level.
Simultaneously, the history information of the underlying replacement policy, if any, is updated.
Please note that ISCA keeps no history information for detecting correlated blocks.

The main operation of ISCA is when a cache miss occurs.
On a cache miss, ISCA conducts the tag comparison operation in the upper set and sends the read request to the lower memory level if the reference is for a read, not a writeback.
If a correlated block is found in the upper set, the victim block is the one located in the set in a way similar to the correlated block.
Otherwise, the lower set is searched as well to find a correlated block.
If not found, a block with minimum distance to its residing way's access port is selected as a victim.
It should be noted that when combined with an underlying replacement policy, ISCA only considers a subset of blocks more suitable for eviction when searching adjacent sets or finding the nearest block to access the port.

After determining the victim block, the data is written (for a read miss, we should wait for data arrival) after shifting the block to the access port, and the history information of the underlying replacement policy and access port pointer are updated.

\section{Simulation Setup and Results}
\subsection{System Setup}
The proposed RTM structure and ISCA replacement policy are evaluated using the gem5 full-system simulator \cite{binkert2011gem5}, modeling a quad-core processor equipped with RTM-based LLC. 
33 combinations of programs from the SPEC CPU2017 benchmark suite \cite{speccpu2017} are used as workloads, and the simulations are performed for 4 billion instructions after fast-forwarding the first 400 million instructions as the warm-up phase.
The RTM timing and energy parameters are extracted from a modified version of DESTINY \cite{mittal2017destiny} and included in gem5 \cite{binkert2011gem5}.
We consider a track length of 16-bit with a single access port. 
The details of system configuration are given in Table \ref{table}.

ISCA is compared with Tape Cache \cite{venkatesan2015cache} with horizontal DBC as the baseline as well as vertical DBC with LRU replacement policy (TapeCache and V-TapeCache hereafter).
The underlying replacement policy for ISCA is LRU with the age threshold of 50\% to only consider the upper half of older blocks as victim candidates. 
We also include a comparison to BlendCache \cite{hameed2022blendcache} and SAR \cite{hameed2023energy} as state-of-the-art RTM-cache management schemes.
Shifts per access (SPA), energy consumption, instructions per cycle (IPC), and miss per kilo instructions (MPKI) are the main evaluated metrics.

\begin{table}[t]
\centering
\caption{System configuration.}\vspace{-5pt}
\includegraphics[width=1\linewidth]{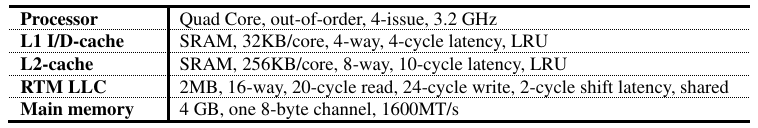}\vspace{-5pt}
\label{table}\vspace{-5pt}
\end{table}

\begin{figure*}[t]
	\centering
	\includegraphics[width=0.85\linewidth]{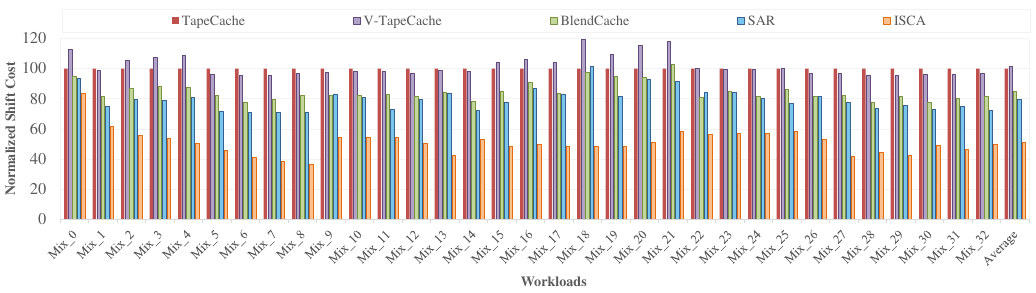}
	\caption{Number of shifts for evaluated schemes normalized to TapeCache baseline for various workloads.}\vspace{-10pt}
	\label {shifts}
\end{figure*}

\subsection{Results}
The main goal of ISCA is to reduce the number of shifts per access in RTM-cache to reduce the cache energy consumption and access latency.
These reductions may be at the cost of increasing cache miss rate due to trading off the performance-oriented existing replacement policy decisions and distance of victim candidates to access ports.
In the following, we first analyze the effect of evaluated schemes on shift operations and energy consumption and then report the IPC and MPKI.

Fig. \ref{shifts} depicts the SPA for all schemes in various workloads. 
While SPA for V-TapeCache is 1.6\% higher than that in the baseline, SAR \cite{hameed2023energy} reduces this value by 20.2\%, and BlendCache \cite{hameed2022blendcache} reduces it by 15.3\% by sacrificing the associativity of the cache and configuring it in a direct-mapped structure.
The SPA in ISCA is reduced by 49.0\%, indicating a 29.7\% and 34.6\% reduction compared to SAR and BlendCache, respectively.
This significant reduction shows high inter-set correlation and ISCA's capability in detecting it among cache blocks.

\begin{figure}[t]
\centering
\includegraphics[width=1\linewidth]{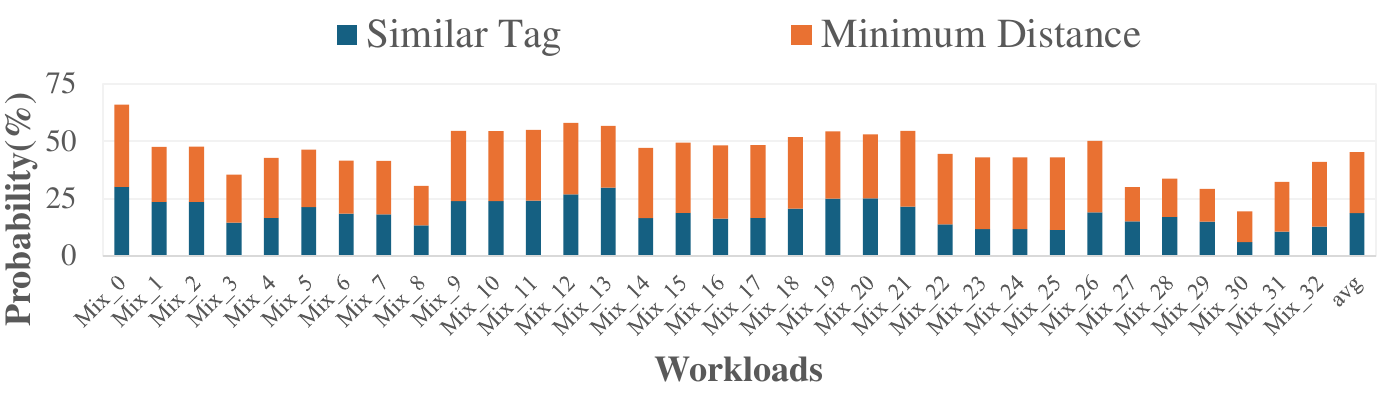}\vspace{-5pt}
\caption{Probability of finding a similar tag in adjacent sets or a block adjacent to one of the access ports on a cache miss in the proposed replacement policy, considering half older blocks in a set.}
\label {SimTagMinDisp}
\end{figure}

Fig. \ref{SimTagMinDisp} illustrates the probability of finding a correlated block in adjacent sets or an adjacent access port by ISCA.
As explained in Section \ref{prop}, the victim block on a cache miss in ISCA is located in a way that a correlated block is found, if any. 
Otherwise, the block in a way with the minimum distance to the access port is selected, considering only the older half of the blocks in the set.
According to Fig. \ref{SimTagMinDisp}, for 18.6\% of cache misses, a correlated block is found for blocks that are old enough for eviction.
For the remaining cache misses, an old block adjacent to the access port is found with the probability of 26.8\%, and the other victim blocks require at least a two-bit shift to be aligned with the access port.

\begin{figure}[t]
\centering
\includegraphics[width=1\linewidth]{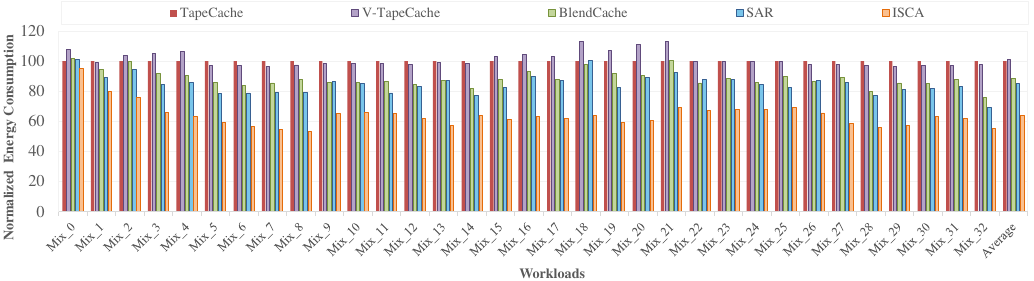}
\caption{Cache energy consumption for evaluated schemes normalized to TapeCache baseline for various workloads.}\vspace{-10pt}
\label {energy}
\end{figure}

Fig. \ref{energy} shows the energy consumption for the evaluated schemes. 
As reported in the recent studies, shift operations contribute more than 50\% of total LLC energy consumption, and the efforts for reduction in the number of shifts are for energy saving.
According to Fig. \ref{energy}, ISCA reduces the cache energy consumption by 35.9\% compared to the H-LRU baseline, while this value is 11.4\% and 15.0\% in BlendCache \cite{hameed2022blendcache} and SAR \cite{hameed2023energy}, respectively.

\begin{figure}[t]
\centering
\includegraphics[width=1\linewidth]{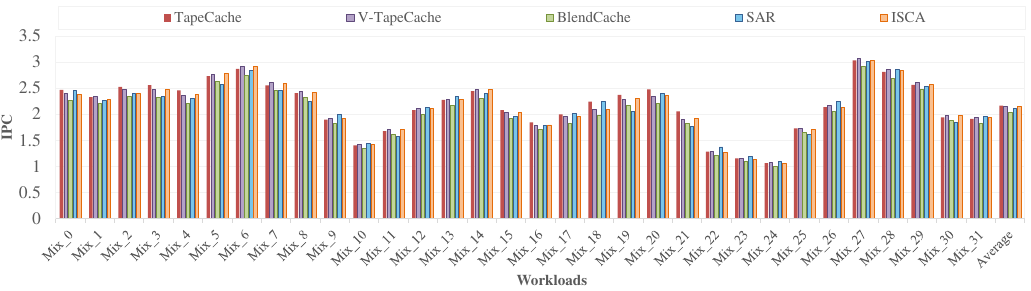}
\caption{Instructions per cycle (IPC) for evaluated schemes normalized to TapeCache baseline for various workloads.}\vspace{-5pt}
\label {ipc}
\end{figure}

Fig. \ref{ipc} shows the IPC of evaluated schemes normalized to H-LRU as the baseline.
Performance degradation in ISCA is the reduction of 1.1\% in IPC, on average.
This reduction for BlendCache \cite{hameed2022blendcache} and SAR \cite{hameed2023energy} is 6.1\% and 2.4\%, respectively.
ISCA, as well as BlendCache \cite{hameed2022blendcache} and SAR \cite{hameed2023energy}, affects the performance in two aspects. Reducing the number of shifts per access enhances the performance by reducing the hit time, and increasing the miss rate degrades the performance.
While for all schemes the latter outperforms the former, the lower overhead of ISCA is the result of higher SPA reduction in the cost of lower miss rate reduction, demonstrating its higher efficiency over the state-of-the-art schemes.

\begin{figure}[t]
\centering
\includegraphics[width=1\linewidth]{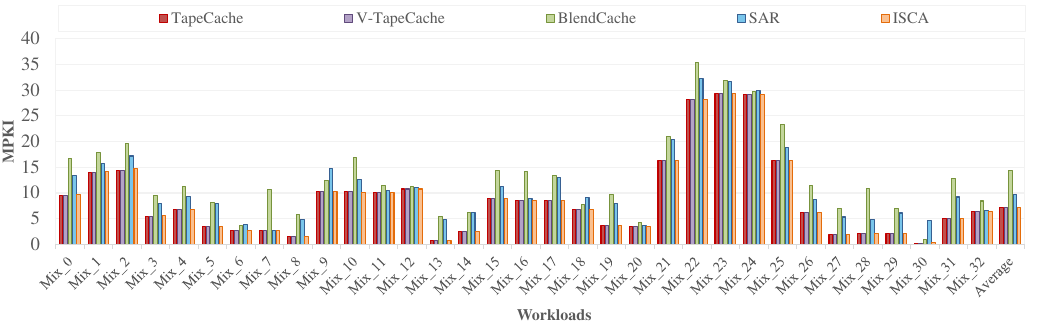}
\caption{Miss per kilo instructions (MPKI) for evaluated schemes normalized to TapeCache baseline for various workloads.}\vspace{-15pt}
\label {mpki}
\end{figure}

Fig. \ref{mpki} depicts the MPKI of the evaluated schemes normalized to the H-LRU baseline.
Trading off between the number of shifts and recency of block accesses in SAR \cite{hameed2023energy} and ISCA and losing associativity in BlendCache \cite{hameed2022blendcache} increase the miss rate.
The results show the superiority of ISCA over SAR \cite{hameed2023energy} and BlendCache \cite{hameed2022blendcache}.
While BlendCache \cite{hameed2022blendcache} cannot compete with the other two schemes due to its direct-mapped structure, the superiority of ISCA over SAR \cite{hameed2023energy} illustrates its higher probability of finding old blocks near the access port.

\subsection{ISCA with Longer Tracks}
The main advantage of RTM technology over other emerging NVM technologies is its significantly higher density, thanks to its unique feature for sharing a single access transistor among tens or even hundreds of ultra-dense data bits.
According to recent reports, the access transistor occupies the majority of the RTM area.
Although longer tracks are preferred in terms of density, increasing the number of shifts and its impact on energy consumption and access latency limits the length of tracks in such a way that some studies included even multiple access ports per track to alleviate this challenge.

Here, we explore the ability of the evaluated schemes (i.e., SAR \cite{hameed2023energy}, BlendCache \cite{hameed2022blendcache}, and ISCA) to extend the length of tracks without unaffordably increasing the number of shifts.
The direct-mapped cache structure of BlendCache \cite{hameed2022blendcache} leads to unrelated blocks residing on a track with high probability because there is no flexibility in cache line allocation that exists in associative caches. 
Therefore, it is expected to significantly increase the shifts by longer tracks in addition to performance degradation imposed by its non-associative structure.
Considering the cache structure in SAR \cite{hameed2023energy}, the track length is architecturally limited to cache associativity as multiple ways in a set share a single track. 
Hence, it is not possible to employ longer tracks without increasing the cache associativity, which complicates the cache and increases the access latency and energy consumption.

\begin{figure}[t]
\centering
\includegraphics[width=1\linewidth]{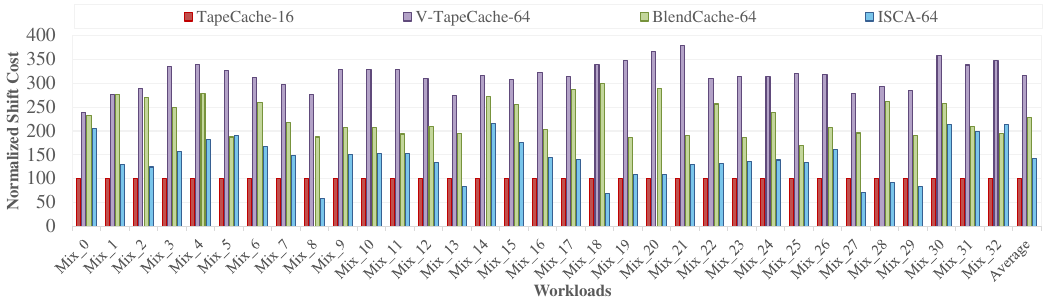}
\caption{Number of shifts considering 64-bit track length (64-set supersets) for evaluated schemes normalized to TapeCache baseline with 16-bit track length for various workloads.}\vspace{-5pt}
\label {shift64}
\end{figure}

ISCA has no architectural limit on the length of tracks, and it can be theoretically extended to the number of cache sets to maximize the cache density.
In terms of performance and energy consumption, however, it should be well-engineered to make a balance among the cache criteria.
Our previous evaluation was based on 16-bit tracks in a 16-way associative cache for a fair comparison.
In the following, we evaluate how longer tracks affect the cache metrics, including the number of shifts, energy consumption, IPC, and area.
The evaluations are conducted for V-TapeCache, BlendCache \cite{hameed2022blendcache}, and ISCA, all with 64-bit tracks, and the baseline is again H-LRU with 16-bit tracks.
Note that 1) we excluded SAR \cite{hameed2023energy} in this evaluation as a 16-way associative cache cannot be constructed by 64-bit tracks in SAR \cite{hameed2023energy} and 2) MPKI in all schemes and the probability of finding correlated blocks in ISCA are not meaningfully affected by increasing the track length and are not reported for the sake of brevity.

Fig. \ref{shift64} depicts the SPA for all schemes in various workloads compared to the baseline. 
By quadrupling the track length, the SPA in V-TapeCache increases from 5.4 to 16.3, on average, indicating a semi-linear trend between SPA and track length.
The increase in SPA for BlendCache \cite{hameed2022blendcache} is from 4.5 in 16-bit tracks to 12.0 in 64-bit tracks, again indicating a semi-linear trend with the track length.
This SPA is 2.3x higher than the baseline, which significantly prolongs the access latency.
The SPA in ISCA is 7.6, which is 1.4x higher than the baseline and 36.6\% and 53.4\% lower than that in BlendCache \cite{hameed2022blendcache} and V-TapeCache.

\begin{figure}[t]
\centering
\includegraphics[width=1\linewidth]{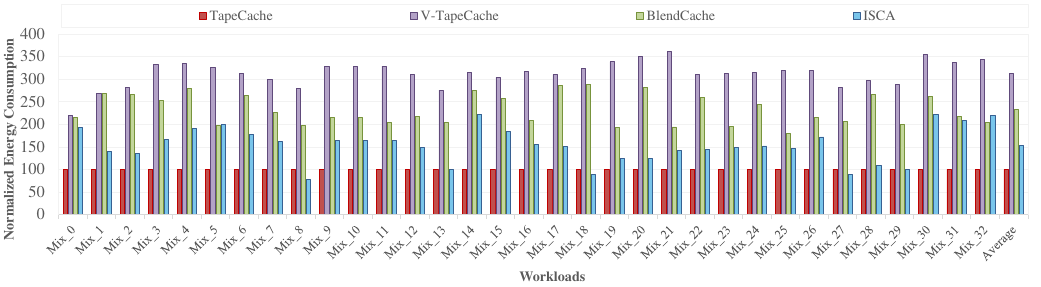}
\caption{Energy consumption considering 64-bit track length (64-set supersets) for evaluated schemes normalized to TapeCache baseline with 16-bit track length for various workloads.}\vspace{-5pt}
\label {threshold}
\end{figure}

Fig. \ref{threshold} shows the cache energy consumption for the evaluated configurations normalized to the baseline. 
By increasing the track length, cache leakage is reduced due to a lower number of access transistors, but access energy increases due to a higher number of shifts.
Our observation illustrates that the cache energy consumption is increased by 212.8\%, 232.0\%, and 54.3\% in V-TapeCache, BlendCache \cite{hameed2022blendcache}, and ISCA, respectively.

\begin{figure}[t]
\centering
\includegraphics[width=1\linewidth]{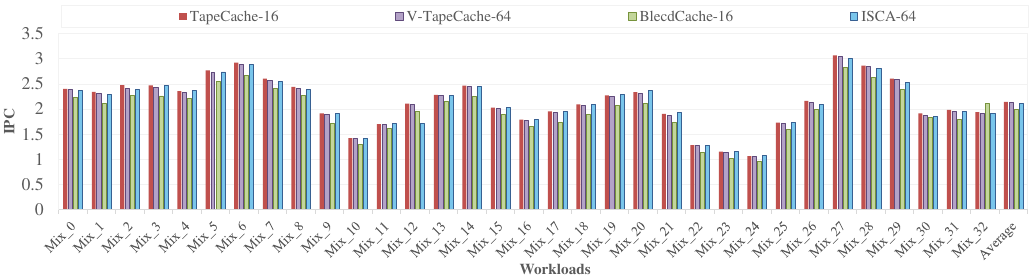}
\caption{Instructions per cycle (IPC) considering 64-bit track length (64-set supersets) for evaluated schemes normalized to TapeCache baseline with 16-bit track length for various workloads.}\vspace{-5pt}
\label {64 tile}
\end{figure}

Fig. \ref{64 tile} shows the effect of longer tracks on the system performance.
Increasing the SPA evidently degrades the performance in favor of higher cache density and area reduction.
The results show that IPC in V-TapeCache, BlendCache \cite{hameed2022blendcache}, and ISCA is lower than the baseline by 1.1\%, 7.7\%, and 1.5\%, respectively, indicating a lower impact of longer track on ISCA compared to the BlendCache configuration.

\subsection{Hardware Complexity and Area Overhead}
The ISCA scheme requires minor hardware modifications with negligible overhead.
The victim selection mechanism requires no history table or metadata storage.
Searching upper and lower sets on a cache miss for finding correlated blocks requires no extra circuitry but a minor modification in cache controlling logic to trigger the tag comparison operation in adjacent sets.
In addition, finding a block with minimum distance to the access port is similar to finding a block with maximum age in LRU replacement policy in terms of hardware complexity and the logic behind required circuitry.

Sharing DBCs among consecutive sets instead of ways inside a set in ISCA has no impact on the cache area but makes it possible to arbitrarily adjust the track length instead of fixing it to the cache associativity.
We have already investigated 64-bit tracks in ISCA and showed that by slightly impacting the IPC and energy consumption, the number of access transistors as the main contributor in the RTM area can be reduced by 4x.
Our evaluation shows that the cache area and static energy consumption are reduced by 43.1\% and 31.7\% when replacing 16-bit tracks with their 64-bit counterpart in ISCA.

\section{Conclusion}
In this work, we have identified critical inefficiencies in conventional data placement strategies for racetrack memory (RTM)-based last-level caches, revealing that redundant shift operations can significantly undermine both performance and energy efficiency. Our novel data placement and replacement scheme addresses these limitations by intelligently grouping correlated blocks on the same track to minimize unnecessary shifts. This approach not only reduces the shift overhead by 49.0\% but also cuts cache energy consumption by 35.9\% in the cost of 1.1\% overall performance degradation, as demonstrated by extensive evaluations. 

These observations open promising new research directions toward realizing the full potential of RTM in high-performance computing systems. Future studies may further explore dynamic data placement strategies, adaptive replacement policies, and the integration of hybrid memory technologies. The proposed scheme represents a significant step toward deploying racetrack-based caches in next-generation processors, offering a pathway to achieve both high performance and energy-efficient computation.

\bibliographystyle{IEEEtran}
\bibliography{sample-base}


\begin{IEEEbiography}[{\includegraphics[width=1in,height=1.25in,clip,keepaspectratio]{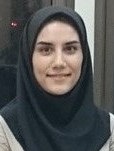}}]{Elham Cheshmikhani}
	is an assistant professor at the Department of Computer Science and Engineering at Shahid Beheshti University, Tehran, Iran. She also served as an assistant professor at the Department of Computer Engineering at Amirkabir University of Technology (Tehran Polytechnic), Tehran, Iran, from September 2022 to September 2023. Her educational background includes a B.Sc., M.Sc., and Ph.D. degrees in Computer Engineering from the Iran University of Science and Technology (IUST), Amirkabir University of Technology (Tehran Polytechnic), and Sharif University of Technology (SUT) respectively, obtained in the years 2011, 2013, and 2020.
	From January 2021 to September 2022, she worked as a postdoctoral researcher at the Department of Information Engineering and Mathematics at the University of Siena, Siena, Italy. Her research interests encompass emerging nonvolatile memory technologies, processing-in-memory, RISC-V ISA design, hardware accelerators, SoC design, dependable systems design, and storage systems.

\end{IEEEbiography}

\vspace{11pt}

\begin{IEEEbiography}[{\includegraphics[width=1in,height=1.25in,clip,keepaspectratio]{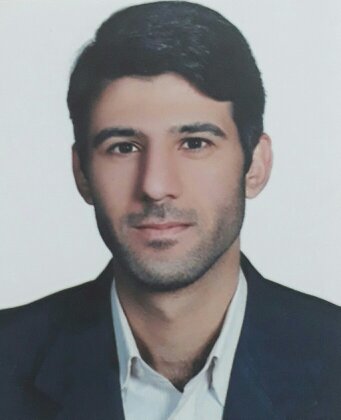}}]{Hamed Farbeh} (S'12--M'18)
	received the B.Sc., M.Sc., and Ph.D. degrees in computer engineering from Sharif University of Technology (SUT), Tehran, Iran, in 2009, 2011, and 2017, respectively. He was a member of the Dependable Systems Laboratory (DSL) at SUT from 2007 to 2017, was with the Embedded Computing Laboratory (ECL), KAIST, Daejeon, South Korea, as a Visiting Researcher from October 2014 to May 2015, and collaborated with the Institute of Research for Fundamental Sciences (IPM), Tehran, Iran, as Postdoc fellow from May 2017 to January 2018. He is currently a faculty member of the Department of Computer Engineering, Amirkabir University of Technology (Tehran Polytechnic-AUT), Tehran, Iran, where he established the Intelligent Computing and Communication Infrastructure Laboratory (ICCI) and is the director of Computer Systems Architecture and Networks (CSAN) group. His current research interests include reliable memory hierarchy, emerging memory technologies, AI processors, and cyber-physical systems.
\end{IEEEbiography}


\end{document}